\documentclass[prl,twocolumn,showpacs,amsmath,amssymb,superscriptaddress]{revtex4}

\usepackage{graphicx}
\usepackage{dcolumn}
\usepackage{bm}
\usepackage{amsmath}

\begin{document}

\preprint{APS/123-QED}

\title{
Graphite--diamond phase coexistence study employing a neural-­network mapping of the \textit{ab initio} potential energy surface
}

\author{Rustam Z. Khaliullin}
\email{rustam@khaliullin.com}
\author{Hagai Eshet}
\affiliation{
Department of Chemistry and Applied Biosciences, ETH Z\"urich, USI Campus, via G. Buffi 13, 6900 Lugano, Switzerland
}
\author{Thomas D. K\"uhne}
\affiliation{
Department of Chemistry and Applied Biosciences, ETH Z\"urich, USI Campus, via G. Buffi 13, 6900 Lugano, Switzerland
}
\affiliation{
Department of Physics and Division of Engineering and Applied Sciences, Harvard University, Cambridge, Massachusetts 02138, USA
}
\author{J\"org Behler}
\affiliation{
Lehrstuhl f\"ur Theoretische Chemie, Ruhr-Universit\"at Bochum, D-44780 Bochum, Germany
}
\author{Michele Parrinello}
\affiliation{
Department of Chemistry and Applied Biosciences, ETH Z\"urich, USI Campus, via G. Buffi 13, 6900 Lugano, Switzerland
}


\date{\today}

\begin{abstract}
An interatomic potential for the diamond and graphite phases of carbon has been created using a neural-­network (NN) representation of the \textit{ab initio} potential energy surface. The NN potential combines the accuracy of a first-principle description of both phases with the efficiency of empirical force fields and allows one to perform, for the first time, a molecular dynamics study, of \textit{ab initio} quality, of the thermodynamics of graphite-diamond coexistence. Good agreement between the experimental and calculated coexistence curves is achieved if nuclear quantum effects are included in the simulation.
\end{abstract}

\maketitle

The ability of carbon atoms to form strong chemical bonds with a variety of coordination numbers leads to a remarkably wide range of physical properties of the condensed phases of carbon. The diamond phase is a three-dimensional network of four-fold coordinated atoms characterized by a very low electrical conductivity and extreme hardness. Unlike diamond, the graphite phase is semimetallic and made up of planes of three-fold coordinated atoms. It behaves as a lubricant because of weak van der Waals (vdW) bonding between the planes.

In spite of the great fundamental and practical importance of graphite and diamond the characterization of these phases and their mutual transformation is far from complete especially in the region of high pressures and temperatures which are difficult to access experimentally. Although computer simulations based on density functional theory (DFT) provide a comprehensive framework for modeling a variety of carbon polymorphs, they become computationally too demanding for the generation of long molecular dynamics (MD) trajectories for large systems (nanosecond-long trajectories are required to study thermodynamics and mechanism of phases transitions). On the other hand, the construction of accurate and computationally efficient potentials capa­ble of describing the wide range of interactions in car­bon is still a challenge. Many simple force fields devel­oped for covalent systems such as the embedded atom method~\cite{a:eamcov}, the Stillinger­-Weber approach~\cite{a:swcarbon}, and the bond-­order potential of Tersoff~\cite{a:tersoff} have only limited suc­cess in modeling carbon phases. More elaborate potentials such as the Brenner potential~\cite{a:brenner}, the environment­-dependent interaction potential~\cite{a:edip}, and a fam­ily of long-­range carbon bond­-order potentials~\cite{a:lcbopii} significantly improve the description of carbon structures by in­corporating $\pi$-­bonding effects and vdW interactions. Nevertheless, even the most sophisticated empirical potentials do not always give a correct description of all properties or phenomena of interest.

In the present paper, we followed a different approach for modeling solid phases of carbon such as diamond and graphite. Instead of representing the interatomic interaction energy by an analytic function fitted to experimental (or \textit{ab initio}) data we created an accurate mapping of the relevant portion of the \textit{ab initio} potential energy sur­face (PES) using a recently developed high-dimensional neural network (NN) approach~\cite{a:behler}. This approach eliminates the requirement to guess a complicated functional form for the in­teratomic potential. Accurate mapping ensures that \emph{all} properties determined by the topology of the PES are described with an accuracy comparable with that of DFT. Furthermore, PES mapping allows one to examine nuclear quantum effects in MD simulations from first principles whereas empirical potentials attempt to incorporate such effects through parameterization. From a computational standpoint, the NN energies, forces, and stress tensor are evaluated with the speed of empirical potentials~\cite{a:behler2,a:behler3} thus enabling an MD study of graphite--diamond coexistence of unprecedented accuracy. 

Neural networks have been successfully used to in­terpolate the PES of simple chemical systems for the last
decade~\cite{a:nnlorenz,a:nnscheffler,a:nnmanzhos,a:nnprudente,a:nnraff1}. However, an NN-­based method that can be used to map the high-dimensional PES of bulk systems and large clus­ters has been introduced only recently~\cite{a:behler,a:behler2,a:behler3}. This mapping of the \textit{ab initio} PES is performed by optimizing NN param­eters to reproduce the \textit{ab initio} energies of many thousands of structures in a training set. The overfitting (i.e. obtaining a good fit to the training data, but performing less accurately when making predictions) is controlled by testing the performance of the NN for an independent test set not used in the optimization.

The accuracy of the reference \textit{ab initio} energies is of paramount importance while training the network. It is known that conventional local- and semilocal density functionals cannot de­scribe the long-­range electron correlations that are respon­sible for the vdW interactions between graphite sheets~\cite{a:marzari}. To account for the dispersion forces in graphite, we employed the Perdew-Burke-Ernzerhof (PBE) functional in combination with the dispersion corrected atom centered pseudopotential (DCACP)~\cite{a:dcacp}, which has been shown to perform well for graphene sheets and aromatic com­pounds~\cite{a:dcacp,a:ursula1}. Extensive tests were performed to demon­strate that DCACP closely reproduces the experimental lat­tice constants as well as elastic and vibrational properties of diamond and graphite (TABLE~\ref{tab:zerok}). The ABINIT package~\cite{a:abinit} was used to perform the \textit{ab initio} calculations. A dense mesh of $k$-­points and a large plane-wave cutoff of 170~Ry were used for all structures so as to ensure convergence of the total energy to 1~meV/atom.

\begin{table*}
\begin{minipage}{\textwidth}
\caption{\label{tab:zerok}Structural, elastic, and vibrational properties of graphite and diamond.}
\begin{ruledtabular}
\begin{tabular}{lcccccccccc}
 &\multicolumn{2}{c}{Lattice const. (\AA)}& \multicolumn{6}{c}{Elastic constants (GPa)} & \multicolumn{2}{c}{Freq. (cm$^{-1}$)}\\
\hline
Hex. graphite         & $a_0$ & $c_0$ & $B_0$ & $c_{11}$ & $c_{12}$ & $c_{33}$ & $c_{44}$ & $c_{13}$ & $\Gamma_{ZO}$ & $\Gamma_{LO/TO}$ \\
\hline
PBE\footnotemark[1]   & 2.461 & 8.712 & 2.4 & \multicolumn{2}{c}{1240\footnotemark[1]} & 2.4 & & -­0.5 & 1561, 1561 & 881  \\
PBE, DCACP            & 2.467 & 6.815 & 37  & 1069 & 162 & 40 & 5 & -­4 & 1553, 1573 & 870  \\
NN                    & 2.467 & 6.688 & 48  & 1080 & 179 & 52 & 7 & 0 & 1527, 1530 & 834  \\
Exp.\footnotemark[2]  & 2.461 & 6.705 & 36.4$\pm$1.1 & 1060$\pm$16 & 180$\pm$20 & 36.5$\pm$1 & 4.0$\pm$0.4 & 15$\pm$5 & 1575 & 861 \\
\hline
Cub. diamond          & $a_0$ &  & $B_0$ & $c_{11}$ & $c_{12}$ & $c_{44}$ &  &  & $\Gamma_{O}$ &  \\
\hline
PBE\footnotemark[1]   & 3.568 &  & 432 & 1060 & 125 & 562 &  & & 1289 &  \\
PBE, DCACP            & 3.570 &  & 439 & 1056 & 130 & 567 &  & & 1292 &  \\
NN                    & 3.569 &  & 434 & 1016 & 142 & 580 &  & & 1295 &  \\
Exp.\footnotemark[3]  & 3.567 &  & 442 & 1076.4$\pm$0.2 & 125.2$\pm$2.3 & 577.4$\pm$1.4 &  &  & 1332 &  \\
\end{tabular}
\end{ruledtabular}
\footnotetext[1]{Results of calculations with the standard Vanderbilt ultrasoft PP from Ref.~\onlinecite{a:marzari}, $c_{11} + c_{12}$ value from Ref.~\onlinecite{a:marzari}.}
\footnotetext[2]{Lattice constants from Ref.~\onlinecite{a:r22}, elastic constants from Ref.~\onlinecite{a:r11}, vibrational frequencies from Refs.~\onlinecite{a:r59,a:r55}.}
\footnotetext[3]{Lattice constants from Ref.~\onlinecite{b:donohue}, elastic constants from Ref.~\onlinecite{a:r50}, vibrational frequency from Ref.~\onlinecite{a:r9}.}
\end{minipage}
\end{table*}

The initial fitting of the carbon NN potential was performed on crystal structures that included the zero­-temperature and randomly distorted structures of cu­bic and hexagonal diamond, hexagonal and rhombohe­dral graphite in the pressure range from -­10 to 200~GPa. After the initial training, the NN was improved self­-consistently by iterative repetition of the NN-­driven MD simulations, collection of new structures emerging from the simulations, calculation of the DFT energies for
the physically relevant structures, and refinement of the NN. These iterations were performed until the root mean
squared error (RMSE) of the new structures not included in the fit converged to the RMSE of the test set. After the self-­consistent procedure the DFT dataset contained $\sim$60,000 DFT energies corresponding to more than 700,000 atomic environments. 10\% of all structures were ran­domly chosen for the test set. The best fit was ob­tained for a NN with 2 hidden layers, each of which contains 25 nodes (the total number of the NN parameters is 1901). The RMSE of the training set is 4.0~meV/atom, while the RMSE of the test set is 4.9~meV/atom.
The maximum absolute errors are 41.5~meV/atom and 46.7~meV/atom for the training and test sets, respectively. The largest errors are attributed to highly distorted graphite structures that are accessible only at temperatures of 4000--5000~K. At these high temperatures the errors are small compared to $k_B T\sim 340-430$~meV so the quality of the relevant ensemble averages is essentially maintained. 

\begin{figure}
\includegraphics*[width=8.5cm]{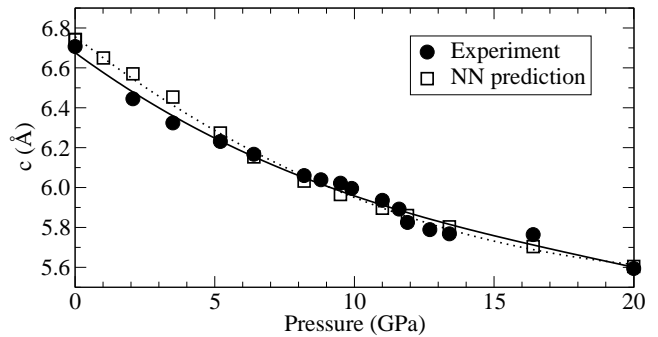}
\caption{\label{fig:cvsp} NN prediction and experimental data~\cite{a:zhao} for the $c_0$ lattice parameter of hexagonal graphite as a function of pressure. $T = 295~K$. The NN values are computed from constant-­pressure MD simulations with a quantum Langevin thermostat~\cite{a:quantumtherm}.}
\end{figure}

To check the accuracy of the NN potential we cal­culated lattice constants, stiffness coefficients, and vi­brational frequencies for the  zero-temperature structures of cubic diamond and hexagonal graphite. The lattice constants were determined by minimization of the NN potential energy fitted using the Murnaghan equation~\cite{a:murnaghan} in the case of diamond and by a two-dimensional fourth order polynomial in the case of graphite. The second­-order elastic constants were calculated by fitting the energy as a function of an appropriate cell distortion to a parabola~\cite{a:fast,a:guo} while vibra­tional frequencies were obtained by diagonalizing the dynamical matrix. The computed quantities are summa­rized and compared with DFT and experimental values in TABLE~\ref{tab:zerok}. The NN accurately reproduces DFT results for structural, elas­tic and vibrational properties of diamond~\footnote{To demonstrate the accuracy of the NN potential relative to DFT, it is sufficient to calculate all properties in TABLE~\ref{tab:zerok} classically without ZPE correction. However, it is worth pointing out that the ZPE corrected PBE value of $a_0$ for diamond is 3.584\AA, significantly larger than the experimental lattice constant of 3.568\AA.}. All properties of graphite determined by the strong in-­plane interactions ($a_0$, $c_{11}$, $c_{12}$, $\Gamma_{ZO}$) are also described well by the NN. However, the relative error between DFT and NN values is generally larger for the properties determined by weak interplanar interactions (e.g. $c_{33}$, $c_{44}$, $c_{13}$). Nevertheless, the NN description of one of the most important structural characteristics of graphite -- the interlayer distance -- is remarkably accurate for a wide range of pressures (FIG.~\ref{fig:cvsp}).

The graphite-­diamond coexistence line was determined by locating points of equal chemical potential in the P--T plane. This was done in three steps. First, we calculated the Helmholz free energy $F^{\text{NN}}(T_0,\rho_0)$ of both phases at $T_0 = 2000$~K by thermodynamic inte­gration using Einstein crystals as the reference systems~\cite{b:frenkel}
\begin{equation}
\label{eq:ti}
F^{\text{NN}}(T_0 , \rho_0 ) = F^{\text{EIN}}(T_0 ,\rho_0 ) + \int_0^1 \langle \frac{\partial U_{\lambda}}{\partial \lambda} \rangle_{\lambda} d\lambda,
\end{equation}
where $U_{\lambda} = \lambda U^{\text{NN}} + (1 - \lambda) U^{\text{EIN}}$.

%
%

In the next step, the chemical potentials 
were evaluated by integrating the free energy as a func­tion of density starting from $\rho_0$~\cite{a:ghir}
\begin{eqnarray}
\label{eq:mu}
\mu^{\text{NN}}(T_0 , \rho) &=& \frac{1}{N} F^{\text{NN}}(T_0, \rho_0) + \frac{a(T_0)}{\rho_0} + b(T_0) \ln \frac{\rho}{\rho_0} \nonumber \\
&+& b(T_0) + c(T_0)(2\rho - \rho_0 ).
\end{eqnarray}
Parameters $a(T_0)$, $b(T_0)$, and $c(T_0)$ were determined by fit­ting the pressure dependence on density using
\begin{equation}
\label{eq:pfit}
P(T_0,\rho) = a(T_0) + b(T_0)\rho + c(T_0)\rho^2.
\end{equation}

Finally, the coexistence line was traced by integrating the Clausius-Clapeyron equation $\frac{dP}{dT} = \frac{\langle \Delta H\rangle}{T \langle \Delta V \rangle}$ using the predictor-corrector scheme of Kofke~\cite{a:kofke}.

It is important to emphasize that long MD trajectories are essential to obtain statistically accurate results for all three steps. Furthermore, it is desirable to perform simulations using large systems as finite size ef­fects can introduce significant errors to the free energies evaluated by thermodynamic integration~\cite{a:frenkel1}. Hence, di­rect \textit{ab initio} MD simulations for large systems (especially with a large plane wave cutoff and a dense $k$-­point mesh) are computationally very demanding for the evaluation of free energies, whereas, the NN provides an affordable and accurate method to determine the coexistence line.

NN-­driven MD simulations were performed for 512 atoms of cubic diamond (cubic box, $\rho_0 = 173.94~\text{nm}^{-3}$) and 960 atoms of hexagonal graphite (4 layers, cell size ratio 2.024:2.104:1, $\rho_0 = 120.02~\text{nm}^{-3}$). The temperature was controlled using a colored­-noise Langevin thermostat that was tuned to provide the optimum sampling efficiency over all relevant vibrational modes~\cite{a:ceriotti}. The time step was set to 0.7~fs. The integral in Eq.~\ref{eq:ti} was evaluated numerically by the Gauss-­Legendre quadrature with 20 points. At each value of $\lambda$, the average value of the integrand and its statistical error were obtained from a 133~ps trajecto­ry.
State points along the 2000~K isotherm were obtained from NPT simulations governed by Nos\'e--Hoover equations of motion with Langevin noise on the particle and cell velocities~\cite{a:feller,a:ceriotti}. Averaging over a 95~ps trajectory was performed for each state point. 
The predictor-corrector algorithm was iterated until pressure had converged to less than 0.05~GPa that required 2-3 iterations of 50~ps each. The total simulation time required to obtain the coexistence line totals $\sim$5~ns for each phase clearly demonstrating the advantage of the NN approach in comparison with the direct \textit{ab initio} simulation.

We performed two separate calculations of the coexistence line. In the first simulation, the Langevin thermostat was tuned to reproduce quantum-mechanical behavior of carbon nuclei using a recently published method of Ceriotti \emph{et al.}~\cite{a:quantumtherm} In the second simulation, the thermostat was fitted to obtain classical behavior of the nuclei. 

\begin{figure}
\includegraphics*[width=8.5cm]{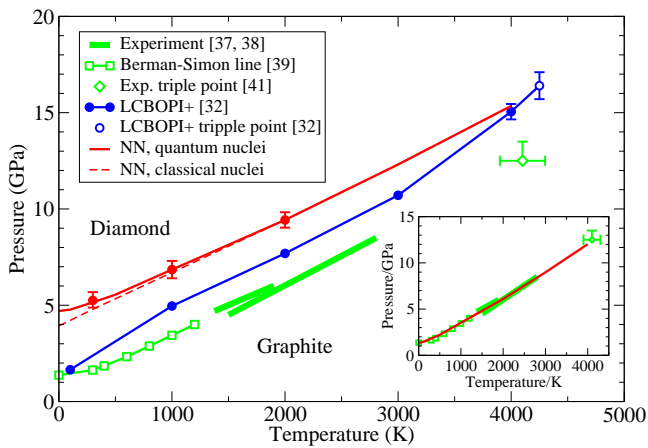}
\caption{\label{fig:phd} (color) Graphite-diamond coexistence line. NN results are denoted by red, LCBOP+ data by blue, and experimental data by green color, respectively.}
\end{figure}

Two graphite-­diamond coexistence lines determined as the intersection of the $\mu^{\text{NN}}(T, P(\rho))$ planes in classical and quantum simulations are shown in FIG.~\ref{fig:phd}. We verified that the coexis­tence lines are calculated correctly by independent thermodynamic integration at $T_0 = 300$~K and $T_0 = 1000$~K (indicated by red points in FIG.~\ref{fig:phd}). Comparison with the experimental data~\cite{a:bundy1961,a:kennedy1976} in the tem­perature interval from 1500 to 3000~K reveals that the NN overestimates the transition pressure by approximately 3.5~GPa. Nevertheless, the slope of the calculated coex­istence line  (2.8$\times$10$^6$~Pa~K$^{-1}$) agrees very well with the ex­perimental value (2.7--3.1$\times$10$^6$~Pa~K$^{-1}$)~\cite{a:bundy1961,a:kennedy1976}.

At temperatures below 1000~K, the quantum coexistence curve flattens out and deviates from the straight classical line (FIG.~\ref{fig:phd}). At 0~K, the quantum transition pressure is 0.8~GPa higher than the corresponding classical value. Analysis of our data shows that this shift is a direct consequence of the diamond zero-point energy being larger than that of graphite. The shape of the calculated quantum coexistence line closely resembles the shape of the Berman-Simon curve obtained from experimental thermodynamic properties of diamond and graphite~\cite{a:berman1955,a:bundy1961}. 
 The 0~K transition pressure predicted by both the NN and PBE functional (4.7~GPa) is again overestimated by 3.3~GPa relative to the experimental value (1.4~GPa)~\cite{a:berman1955,a:bundy1961}. Based on this observation we infer that the positive 3.3~GPa shift of the calculated co­existence line is caused by inaccuracies of the \textit{ab initio} PES and not by errors in the NN mapping. This systematic shift is a result of the inability of the PBE functional to capture precisely the small difference between the energies of graphite and diamond (i.e. $\Delta U^{PBE}_{d-g} = 68$~meV/atom is smaller than the average error of the PBE functional, $160$~meV/atom~\cite{a:pbeerror}). Despite this error, the PBE functional and NN predict the zero-point energy contributions for diamond and graphite correctly and, therefore, accurately describe the flattening of the coexistence line at the low temperatures. The inset of FIG.~\ref{fig:phd} shows that the Berman-Simon curve~\cite{a:berman1955}, the coexistence line of Bundy~\cite{a:bundy1961} and the experimental estimate of the graphite-diamond-liquid triple point~\cite{a:bundy1963} are well reproduced in our calculations if the quantum NN curve is shifted down by 3.3~GPa to match the experimental 0~K transition pressure.

In the 1000--3000~K range, the coexistence line predicted with the LCPOBI+ potential~\cite{a:ghir} lies $\sim$2~GPa closer to the experimental line than the NN curve. However, the LCBOPI+ potential incorrectly predicts an increase in the slope of the line below 1000~K and above 3000~K. As a consequence, the LCBOPI+ triple point lies $\sim$4~GPa above the experimental value even though the 0~K transition pressure is correctly estimated by LCBOPI+.


In summary, we have demonstrated that despite the distinct nature of bonding in graphite and diamond the newly developed NN potential predicts numerous properties of both phases in quantitative agreement with DFT and experimental data. The computational efficiency of the NN potential enables an MD study of graphite--diamond coexistence of unprecedented accuracy. Comparison of the coexistence lines determined in quantum and classical simulations has shown that nuclear quantum effects are responsible for the experimentally observed flattening of the coexistence curve at temperatures below 1000~K. A detailed MD study of the mechanism of the graphite-to-diamond transformation and refinement of the NN potential so as to include high-pressure solid and liquid phases of carbon are useful follow-on developments of this work.

\begin{acknowledgments}
The authors would like to thank M. Ceriotti and G. Bussi for their help with MD simulations that used a Langevin thermostat and barostat and G. Tribello for carefully reading the manuscript. JB is grateful for financial support by the FCI and the DFG. Our thanks are also due to the Swiss National Supercomputing Centre and High Performance Computing Group of ETH Z\"urich for computer time. 
\end{acknowledgments}

\bibliography{carbon}

\end{document}